# Electron interaction-driven insulating ground state in $Bi_2Se_3$ topological insulators in the two dimensional limit

Minhao Liu[1,*], Cui-Zu Chang[1,2,*], Zuocheng Zhang[1], Yi Zhang[2], Wei Ruan[1], Ke He[2,†], Li-li Wang[2], Xi Chen[1], Jin-Feng Jia[1], Shou-Cheng Zhang[3,4], Qi-Kun Xue[1,2], Xucun Ma[2], and Yayu Wang[1,†]

[1]*Department of Physics, Tsinghua University, Beijing 100084, P. R. China*

[2]*Institute of Physics, Chinese Academy of Sciences, Beijing 100190, P. R. China*

[3]*Department of Physics, Stanford University, Stanford, California 94305-4045, USA*

[4]*Institute for Advanced Study, Tsinghua University, Beijing 100084, P. R. China*

\* *These authors contributed equally to this work.*

[†] Email: kehe@aphy.iphy.ac.cn; yayuwang@tsinghua.edu.cn

We report a transport study of ultrathin $Bi_2Se_3$ topological insulators with thickness from one quintuple layer to six quintuple layers grown by molecular beam epitaxy. At low temperatures, the film resistance increases logarithmically with decreasing temperature, revealing an insulating ground state. The sharp increase of resistance with magnetic field, however, indicates the existence of weak antilocalization, which should reduce the resistance as temperature decreases. We show that these apparently contradictory behaviors can be understood by considering the electron interaction effect, which plays a crucial role in determining the electronic ground state of topological insulators in the two dimensional limit.

Topological insulators (TI) are a new class of insulators with topologically nontrivial band structures originated from strong spin-orbit coupling [1-3]. The discovery of quantized spin Hall effect (QSHE) in two-dimensional (2D) TI [4, 5] stimulated intensive search for new TI systems and associated novel phenomena. Recently, three dimensional (3D) TIs have been predicted and experimentally verified in a series of compounds exemplified by $Bi_{1-x}Sb_x$, $Bi_2Se_3$, $Bi_2Te_3$, and $Sb_2Te_3$ [6-9]. These materials possess an insulating energy gap in the bulk and gapless surface state (SS) protected by time-reversal symmetry. The topologically protected SSs have been proposed to host a variety of exotic magnetoelectric phenomena [10, 11], and are ideal platforms for spintronic and quantum computational devices.

Many unique features of the TI surface states, such as Dirac-like linear band dispersion [8, 12], chiral spin texture [13], absence of backscattering [14-16], and Landau level quantization [17, 18], have been revealed by surface sensitive probes including angle-resolved photoemission spectroscopy (ARPES) and scanning tunneling microscopy (STM). The transport properties of the SSs, however, are often mixed with those from the bulk states because it is difficult to grow bulk insulating samples. This has been a main hurdle for realizing the proposed novel phenomena and applications of TIs. Significant reduction of bulk carriers can be achieved by doping/annealing [19-21] of bulk crystals and gate tuning of nanostructures [22, 23]. Another effective approach is to fabricate ultrathin TI films or ribbons to enhance the surface to volume ratio [24]. This approach not only reduces the bulk conduction, but also allows us to tune the coupling of the two surfaces to reach the 2D limit of TIs [25, 26]. Although TIs in the 2D limit have been proposed to exhibit QSHE [25] and enhanced thermoelectric performance [27], transport properties in this regime have not been investigated in a systematic manner.

In this Letter we report electrical transport studies on $Bi_2Se_3$ films with controlled layer thickness from 1 quintuple layer (QL) to 6QL. At sufficiently low temperatures, the film resistance shows a logarithmic increase with reducing temperature, indicating an insulating ground state. The weak field magnetoresistance (MR), on the contrary, always shows a positive cusp characteristic of weak antilocalization (WAL). We propose that the existence of the insulating ground state in the presence of WAL is a clear indication of strong electron interaction effect in the 2D limit of TIs.

The ultrathin $Bi_2Se_3$ films studied here are grown on sapphire substrate by molecular beam epitaxy (MBE). The structure and thickness of the films are monitored *in situ* by reflection high energy electron diffraction (RHEED). The quality of the sample can be seen from the STM image of a 4QL film shown in Fig. 1(a). Except for nanoscale 1QL-high voids and plateaus, the film is atomically flat over large scale. After the growth of a $Bi_2Se_3$ film, 20nm of highly insulating amorphous Se layer is deposited on top of it to prevent the film from direct exposure to air. Ti/Au electrodes are then deposited on the sample, forming millimeter sized transport devices as schematically shown in Fig. 1(b). The total 2D carrier density estimated from the Hall effect measurement is around $(3.5\pm0.5)\times10^{13}/cm^2$, depending on film thickness. The SS carrier density estimated from the ARPES measured Fermi wavevector ($k_F \sim 0.1 Å^{-1}$) [28] is around $1.6\times10^{13}/cm^2$ for each surface. The dominance of SS charge transport is thus justified.

Figure1(c) shows the 2D sheet resistance ($R_\square$) vs. temperature (*T*) for ultrathin $Bi_2Se_3$ films with thicknesses *d* = 1QL to 6QL. The 1QL film is highly insulating with $R_\square$ much larger than $h/e^2$, the quantum resistance. This is most likely due to the poor interface between the first Se layer and the sapphire substrate caused by the large lattice mismatch. As $d$ is increased to 2QL, the $R_\square$ value drops quickly to a fraction of $h/e^2$, indicating much improved film quality. As $d$ increases further, $R_\square$ keeps dropping, mainly due to the enhanced electron mobility as the lattice relaxes continuously. The mobility at $T = 2K$ is estimated to be 31cm$^2$/V.s for the 2QL film and 350cm$^2$/V.s for the 6QL film. For all the films between 2QL and 6QL, the $R_\square$ vs. $T$ curves show metallic (with positive slopes) behavior at high temperatures. At sufficiently low temperatures, however, the $R_\square$ values reach a minimum and then increase with decreasing $T$, indicating the existence of an insulating ground state.

To elucidate the origin of the resistance upturn at low $T$, we renormalize the temperature dependence of $R_\square$ by the minimum resistance and display it in logarithmic scale in Fig. 2. As indicated by the broken line, the $R_\square$ of the 2QL film increases logarithmically at low $T$. As the film thickness increases, the same behavior persists, although the slope of the logarithmic upturn keeps decreasing. Moreover, the $T$ at which $R_\square$ reaches the minimum, designated as $T_{\min}$ hereafter, also decreases continuously as the film thickness increases. Both trends indicate that the insulating tendency is much stronger in thinner films. The inset of Fig. 2 summarizes the $T_{\min}$ of all the samples. A monotonic increase of $T_{min}$ with reducing thickness is clearly demonstrated.

Figure 3(a) displays the normalized MR of different films measured at $T = 1.5K$ in a perpendicular magnetic field. The overall pattern of the MR curves evolves systematically with film thickness. In the 2QL film, MR shows a steep increase at low field and starts to saturate at around 10 T. The magnitude of MR drops considerably in the 3QL film, but the qualitative behavior remains the same. As *d* is further increased to above 4QL, the zero field cusp is still present, but confined to a progressively narrower field scale. In the high field regime the MR evolves into a parabolic *H* dependence. The parabolic contribution becomes larger in thicker films, reflecting the increased weight of bulk MR, which is known to be large, positive, and parabolic to *H*. The wiggles on the MR curves are not noises, but retraceable signals reminiscent of the conductance fluctuations found in $Bi_2Se_3$ single crystals [19].

The sharp increase of weak field MR is a characteristic behavior of weak antilocalization, as has been seen in single crystals [19], nanoribbons [24], and thin films [22, 29] of TIs. Due to the Berry phase associated with the chiral spin texture, the SS electrons travelling along two time-reversed self-intersecting loops accumulate a π phase difference. The destructive quantum interference between them reduces the backscattering probability, which in turn leads to a decrease in resistivity. When an external magnetic field is applied perpendicular to the film, the enclosed flux causes a phase shift between the two loops. Therefore the WAL is suppressed by *H*, leading to a positive MR that can be described by the Hikami-Larkin-Nagaoka (HLN) theory [30]. The HLN formula for weak field conductance variation in the 2D limit is:

$$\delta\sigma = \sigma(B) - \sigma(0) = -\frac{\alpha e^2}{2\pi^2 \hbar}[\ln(\frac{\hbar}{4eBl_\phi^2}) - \psi(\frac{1}{2} + \frac{\hbar}{4eBl_\phi^2})],$$

where $l_\phi$ is the phase coherence length, $\psi$ is the digamma function, and $\alpha$ is a coefficient equals to −1/2 in the symplectic case when WAL dominates.

Shown in Fig. 3(b) are theoretical fits to the low field magnetoconductance ($\delta\sigma$) of the 2QL, 3QL, 5QL, and 6QL films measured at *T* = 1.5 K. The HLN formula gives an excellent fit to all the curves, which unambiguously demonstrates the existence of WAL in the 2D limit of TIs. One systematic trend revealed here is the larger field scales, hence the shorter phase coherence length, in thinner films. Fig. 3(c) summarizes the thickness dependence of the $l_\phi$ value extracted from the HLN fit. It increases from around 75nm in 2QL film to more than 200nm in the 6QL film. The $\alpha$ value ranges between -0.3 and -0.6, in rough agreement with the symplectic case.

Although WAL gives an excellent description of the MR cusp, it also implies that the resistance should decrease logarithmically with lowering $T$, which is contradictory to the observed logarithmic increase of $R_\square$. We propose that the most plausible explanation for these apparently contradictory results is the strong Coulomb interaction between electrons in ultrathin TIs. As first realized by Altshuler and Aronov (AA) [31], the Coulomb repulsion in a 2D disordered metal is retarded due to the diffusive motion of charge carriers. Electron-electron interactions cannot be screened immediately and thus become strong and long-ranged. The enhanced interaction causes a suppression of electron density of state at the Fermi level and a logarithmic increase of resistivity with reducing $T$. When the AA interaction effect and WAL coexist, they give opposite correction to resistivity. The nature of the electronic ground state depends on the relative strength of the two competing effects.

In the $Bi_2Se_3$ films studied here, the interaction effect is enhanced in thinner films due to reduced dimensionality and increased disorderness, both leading to poorer screening. The WAL effect, on the contrary, is greatly weakened as the film becomes thinner. This is opposite to the usual expectation of enhanced WAL as the system becomes more 2D with larger probability of forming self-intersecting loops, but is unique to the TIs in the 2D limit.

As the thickness of Bi$_2$Se$_3$ is reduced to $d < 6$QL, the wavefunctions of the top and bottom SSs start to hybridize, leading to the opening of an energy gap at the Dirac point [26]. The topological protection of the SSs, hence the WAL effect, is weakened by this coupling. As discussed by a recent theory [27], the WAL correction to conductance can be expressed as $\delta\sigma/\sigma = -\cos\delta\phi$, where $\delta\phi$ is the phase difference between two opposite loops. In the 2D regime of TIs with coupled SSs, $\delta\phi$ can be expressed as $\delta\phi = \frac{\varepsilon^2 - \Delta_f^2}{\varepsilon^2}\pi$. Here $\varepsilon$ is the energy of the SS electrons relative to the Dirac point and $\Delta_f$ is the gap amplitude. Using the ARPES spectrum measured on the same films studied here [28], we found that $\delta\phi$ decreases from $\pi$ in the 6QL film to around $0.6\pi$ in the 2QL film, mainly due to the increase of $\Delta_f$. This implies that the relative conductance increase due to WAL is reduced from 100% in the 6QL film to merely 30% in the 2QL film. The combined effect of enhanced interaction strength and weakened WAL with reduced thickness gives a qualitative explanation of the stronger insulating tendency in thinner films.

The electron interaction effect also manifests itself in the field dependence of conductivity. In the AA interaction mechanism, spin splitting by external field leads to a decrease of conductivity with $H$. For $h = g\mu_B H/k_B T \gg 1$, i.e., Zeeman splitting much larger than thermal activation, $\delta\sigma$ due to this effect has a logarithmic $H$ dependence as $\delta\sigma_{AA} \sim -\ln(h/1.3)$ [32]. Unlike the $\delta\sigma$ from WAL ($\delta\sigma_{WAL}$), which is an orbital effect only sensitive to perpendicular magnetic field, $\delta\sigma_{AA}$ is isotropic to $H$. Fig. 4(a) shows the $\delta\sigma_\parallel$ measured for $H$ parallel to the 3QL film (red), which has a sizable variation apparently coming from the spin splitting $\delta\sigma_{AA}$ term. For $H$ perpendicular to the film (black), $\delta\sigma_\perp$ consists of both the $\delta\sigma_{WAL}$ term and $\delta\sigma_{AA}$ term. Fig. 4(b) displays $\delta\sigma_\perp - \delta\sigma_\parallel$, the difference between $\delta\sigma_\perp$ and $\delta\sigma_\parallel$, which should come exclusively from the orbital $\delta\sigma_{WAL}$ term. The curve can still be fitted nicely using the HLN formula up to 2T, giving a $l_\phi \sim 144$nm, about 30% larger than the value extracted from fitting the $\delta\sigma_\perp$ data without subtracting the $\delta\sigma_\parallel$ term (Fig. 3(b)).

In summary, transport studies of ultrathin $Bi_2Se_3$ TI with coupled SSs reveal an unconventional insulating ground state where electron interaction plays a crucial role. The intricate interplay between electron interaction and strong disorder in a topologically nontrivial 2D electron system may give rise to a wealth of new phenomena, such as the recently proposed quantum criticality [33], that was not captured in the single particle picture. The systematic transport studies in well-controlled ultrathin $Bi_2Se_3$ films also pave the road for fabricating and investigating sophisticated TI-based electronic devices.

We thank X. Dai, Z. Fang, N. P. Ong, E. Y. Shang, C. S. Tian, and Z. Y. Weng for helpful discussions. This work was supported by the National Natural Science Foundation of China, the Ministry of Science and Technology of China, and the Chinese Academy of Sciences.

**Figure Captions:**

FIG 1. (a) An *in situ* STM image of a $Bi_2Se_3$ film with thickness $d = 4QL$. The film is flat over large length scale, except for nanoscale islands and voids with 1QL thickness. (b) Schematic structure of the $Bi_2Se_3$ ultrathin film for transport measurements (the thickness is not to scale). (c) The systematic evolution of $R_\square$ vs. $T$ curves for $Bi_2Se_3$ with thickness $d =$ 1QL to 6QL.

FIG 2. The normalized $R_\square$ shows a logarithmic increase with decreasing $T$ (the broken lines are guides to the eyes). Both the slope of the logarithmic upturn and the $T_{min}$ increase with reducing $d$, indicating stronger insulating behavior in thinner films. Inset: the monotonic increase of $T_{min}$ with reducing $d$.

FIG 3. (a) The high field MR of the 5 films with $d = 2QL$ to 6QL all measured at $T = 1.5$ K. In the weak $H$ regime the MR shows a steep increase characteristic of WAL. (b) The HLN fit of the weak field $\delta\sigma$ for films with different $d$ measured at $T = 1.5$ K. (c) The variation of $l_\phi$ with film thickness.

FIG 4. (a) The variation of conductance of the 3QL film for field parallel ($\delta\sigma_\parallel$, red) and perpendicular ($\delta\sigma_\perp$, black) to the film measured at $T = 1.5$K. The wiggles in the high field regime of $\delta\sigma_\perp$ are retraceable signals reminiscent of the conductance fluctuations. (b) The difference of $\delta\sigma_\perp$ and $\delta\sigma_\parallel$ and the theoretical fit using the HLN formula.

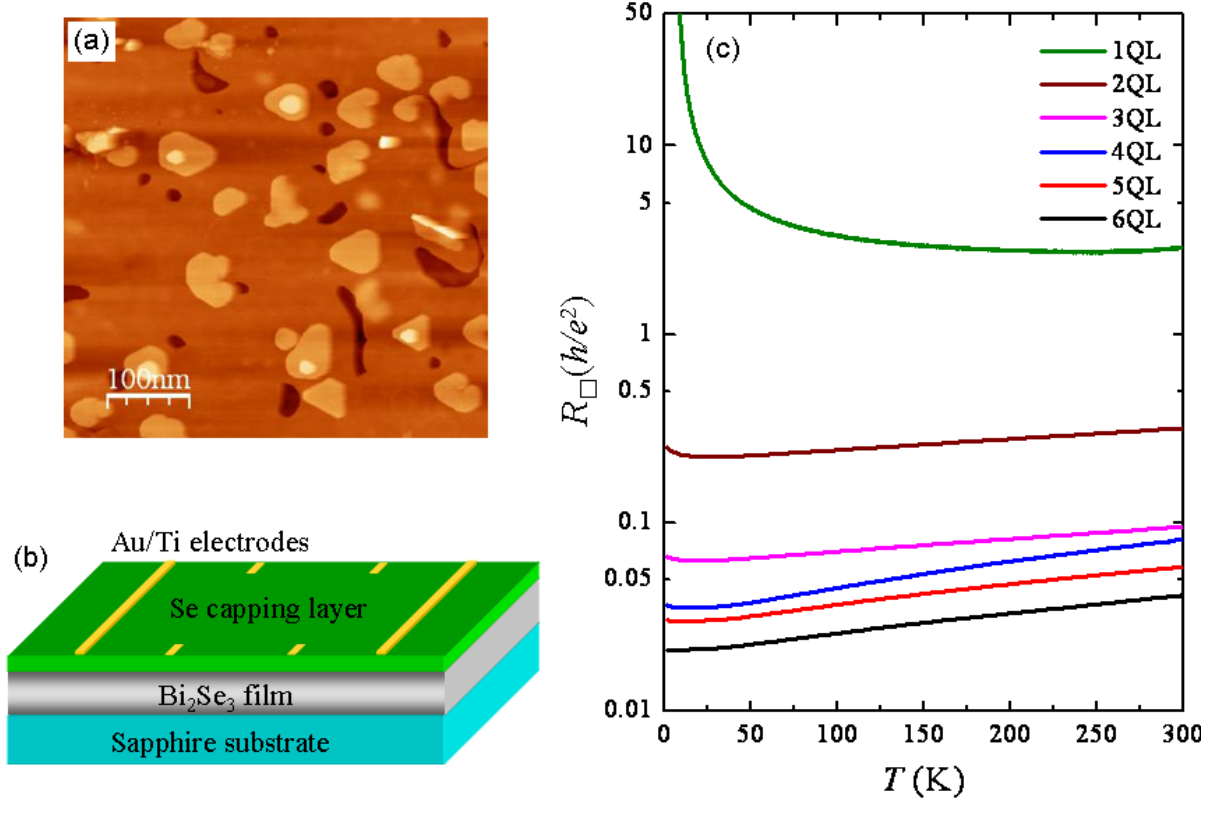

Figure 1

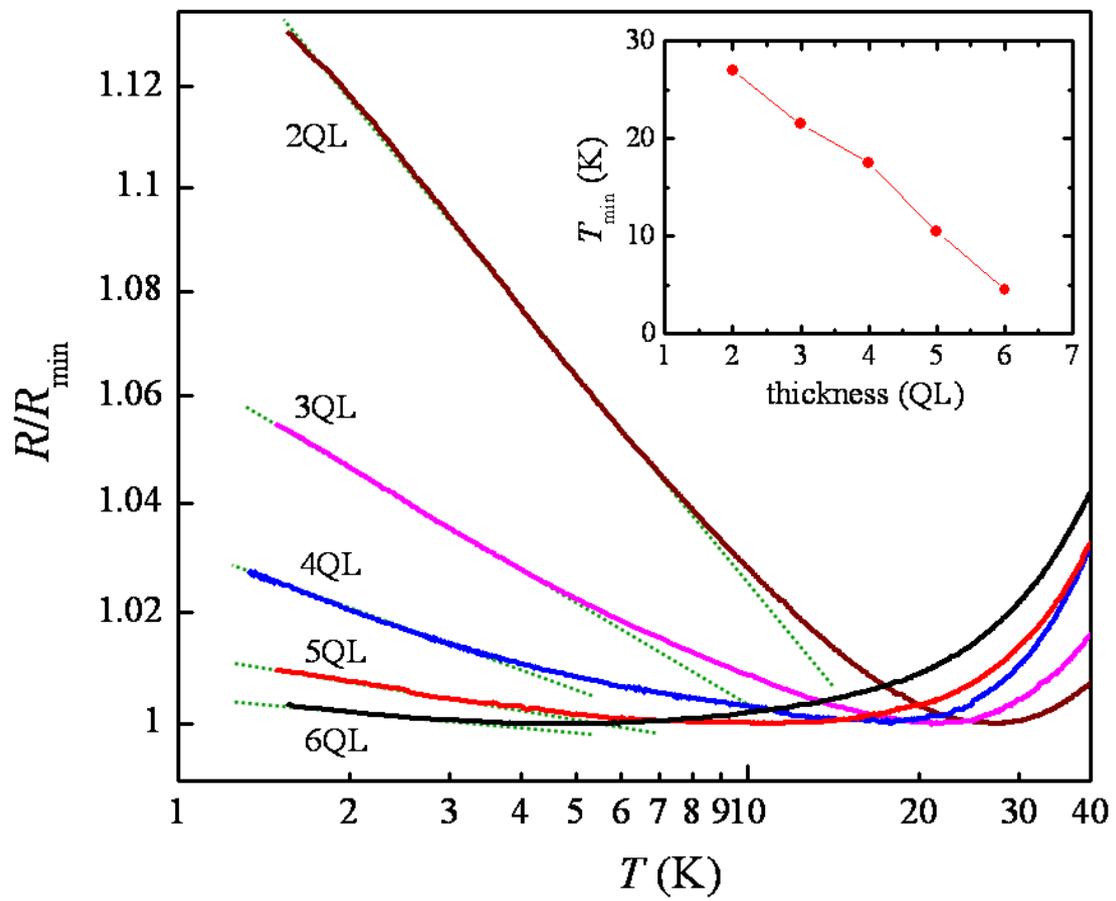

Figure 2

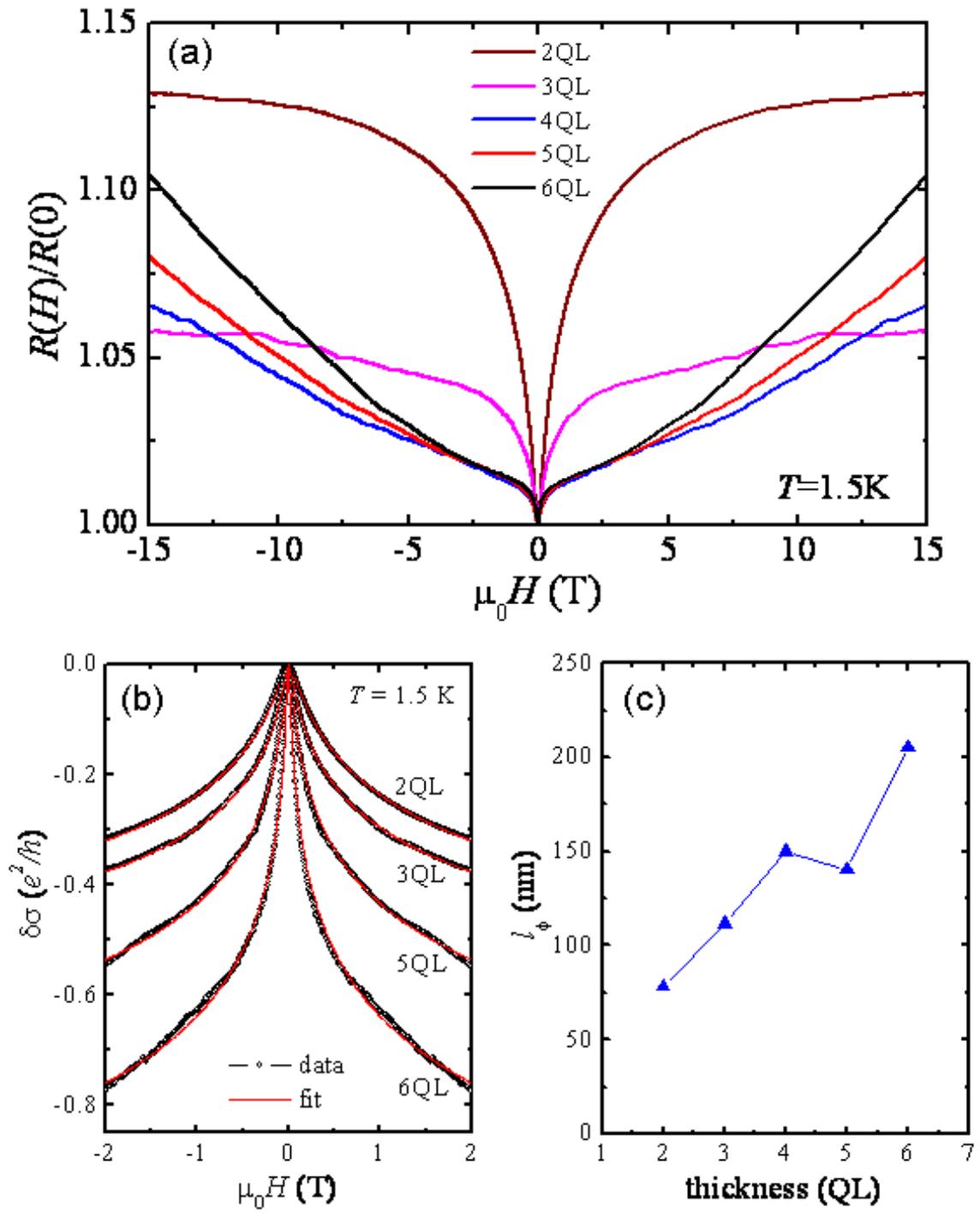

Figure 3

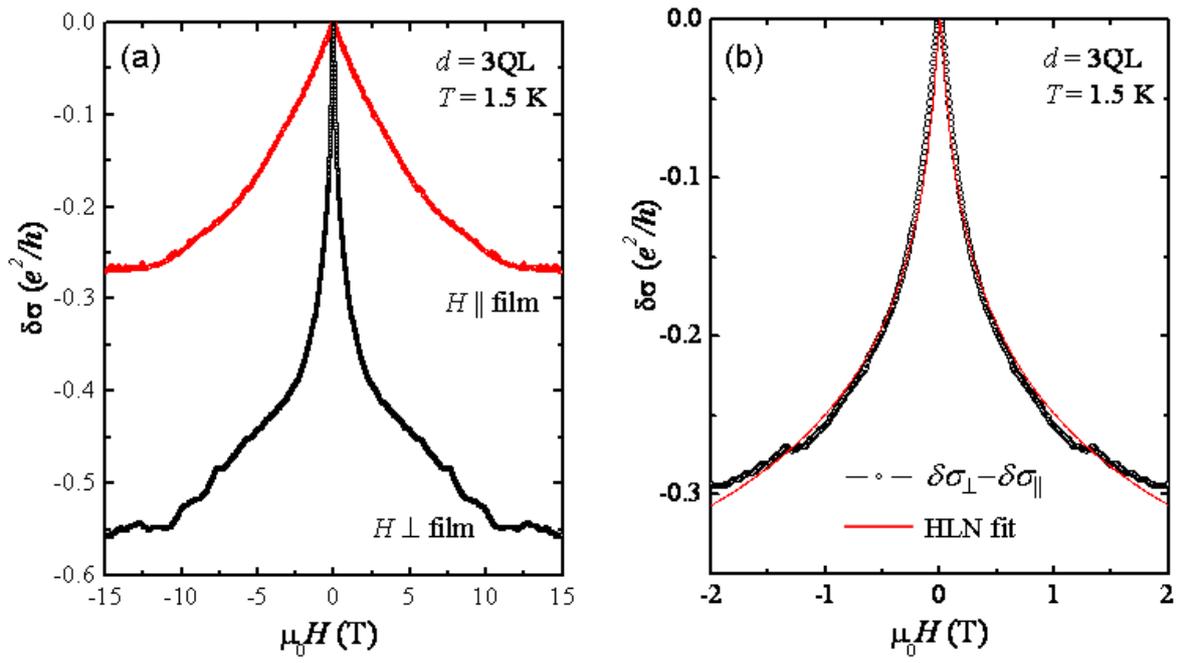

Figure 4